# How reliable are Hanle measurements in metals in a three-terminal geometry?

Oihana Txoperena[1], Marco Gobbi[1], Amilcar Bedoya-Pinto[1], Federico Golmar,[1] Xiangnan Sun[1], Luis E. Hueso[1,2] and Fèlix Casanova[1,2]

[1]CIC nanoGUNE, 20018 Donostia-San Sebastian, Basque Country (Spain)
[2]IKERBASQUE, Basque Foundation for Science, 48011 Bilbao, Basque Country (Spain)

We test the validity of Hanle measurements in three-terminal devices by using aluminum (Al) and gold (Au). The obtained Hanle and inverted Hanle-like curves show an anomalous behavior. First, we measure Hanle signals 8 orders of magnitude larger than those predicted by standard theory. Second, the temperature and voltage dependences of the signal do not match with the tunneling spin polarization of the ferromagnetic contact. Finally, the spin relaxation times obtained with this method are independent of the choice of the metallic channel. These results are not compatible with spin accumulation in the metal. Furthermore, a scaling of the Hanle signal with the interface resistance of the devices suggests that the measured signal is originated in the tunnel junction.

Spintronics is a rapidly growing research area that aims to use and manipulate not only the charge, but also the spin of the electron.[1] A spintronic device capable of creating, transporting, manipulating and detecting spins is a long-sought goal. There is special interest in purely electrical spin injection and detection devices with long, active semiconductor channels, to integrate the spin functionality into conventional electronics. A simple device to study spin injection and transport in semiconductors uses a three-terminal (3T) geometry, in which spin accumulation is induced and probed by a single magnetic tunnel contact.[2,3] The detection is possible because a transverse magnetic field reduces the spin accumulation due to dephasing during the precession, the so-called Hanle effect[3,4]. Since the 3T geometry does not require submicron-sized fabrication processes, this type of devices have become very popular.[2,3,5-16]

However, many of the results associated to this method are still controversial. For example, large disagreements between the theoretically predicted and experimentally measured spin accumulation have been found.[5] Experimental values well above the theoretical ones were first reported in $Fe/Al_2O_3/GaAs$ structures.[6] Since then, similar results have been observed in various semiconductors.[3,5,7-11] Moreover, some other reported features such as an anomalous bias dependence of the Hanle signal[7,8,12,13] or the unclear origin of the inverted Hanle effect[11,14], put the 3T-measurements strongly into question.

In order to gain insight about the reliability of this method, we have applied it to metals with well-known spin transport properties such as aluminum (Al)[4,17-21] and gold (Au).[21-24] In this letter, we report that measurements done with a 3T geometry lead to Hanle- and inverted Hanle-like features which are not compatible with spin accumulation in the studied metals. In addition, the measured signals scale with the interface resistance of the tunnel barrier of the devices, suggesting that the origin of these anomalous signals might arise from the tunnel barrier itself.

We produced devices with different non-magnetic metals: samples 1-3 with the structure Al/AlO$_x$/Py, where the Al (Py) thickness is 15 nm (10 nm), and samples 4-6 with the ferromagnetic metal at the bottom Py/AlO$_x$/Au, with 10 nm of Py and 12 nm of Au. The devices were fabricated in a UHV e-beam evaporation chamber (base pressure <1·10$^{-9}$ mbar), using an integrated shadow masking system. The device geometry is sketched in Fig. 1(a). We used two different strategies for the tunnel barrier fabrication: a plasma exposure to oxidize the Al in a single run (Sample 1); and a three-step deposition of a 6 Å Al layer with subsequent oxidation in an O$_2$ atmosphere without plasma (Samples 2-6). The contact area ranged between 250 x 250 µm$^2$ and 250 x 500 µm$^2$. Electrical measurements were performed by a dc method using a dc current source and a nanovoltmeter,[17] while the temperature and field control is done in a commercial Quantum Design PPMS cryostat.

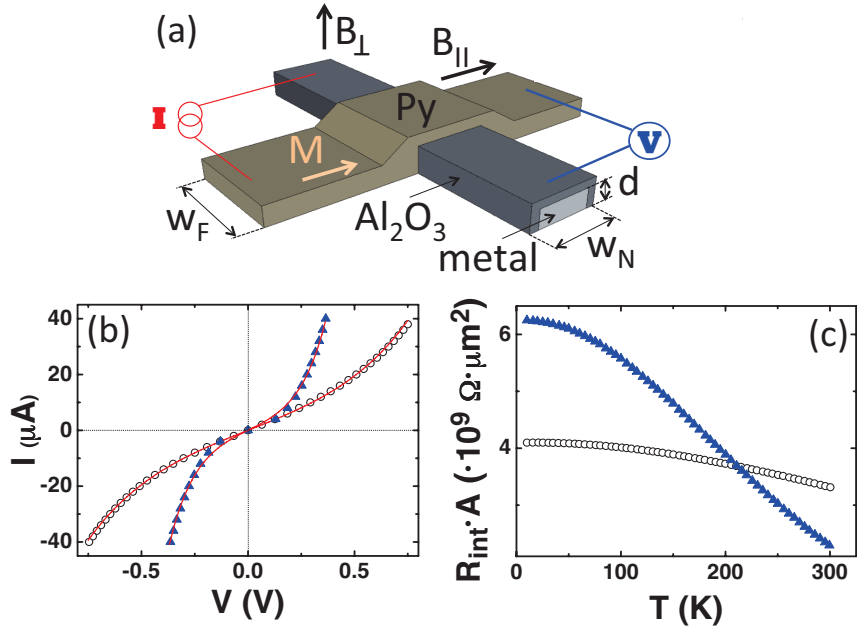

FIG. 1. (a) Scheme and dimensions of the device used for the Hanle measurements. The electrical configuration and the magnetic field direction for both Hanle ($B_\perp$) and inverted Hanle ($B_{II}$) effect are shown. The magnetization vector of Py strip (M), parallel to its easy axis, is also shown. (b) I-V data for Sample 1 (open circles) and Sample 2 (solid triangles), measured at 10 K. Red solid lines are best fits to the standard Simmons model. (c) Zero bias resistance of the tunnel junction, multiplied by its area $A = w_F w_N$, as a function of temperature for Sample 1 (open circles) and Sample 2 (solid triangles).

First, we characterize the tunnel junctions obtained by different fabrication processes. In Fig. 1(b), we show the current-voltage data of the tunnel junctions of Al samples (Samples 1 and 2) measured at 10 K. The measured data is fitted following the standard Simmons model for tunneling.[25] From the fittings (red curves in Fig. 1(b)), we obtain both the height of the tunnel barriers and their thickness: $\phi = 2.8$ V and $d = 1.5$ nm for Sample 1, and $\phi = 0.8$ V and $d = 2.7$ nm in the case of Sample 2. The difference in the barrier parameters is due to the different fabrication strategy of the junctions, as expected.[26,27] Samples 3-6 show barrier parameters similar to Sample 2 (not shown). Fig. 1(c) shows the resistance of the tunnel junction ($R_{int}$) measured at zero bias and multiplied by its total area ($A$) as a function of temperature for Samples 1 and 2. In both cases, we observe a weak decrease with increasing temperature, as expected from a tunnel junction with no pinholes.[27] The values of resistivity of the metal stripes at 10 K, measured with four probes, are 11.3 µΩ·cm (Sample 1, Al), 20.2 µΩ·cm (Sample 2, Al) and 3.5 µΩ·cm (Sample 4, Au).

With the measurement configuration shown in Fig. 1(a), we should observe the Hanle effect when a transverse magnetic field ($B_\perp$) is applied. The spins, accumulated at the interface due to electrical injection from the Py contact into the metal, start to precess around the transverse magnetic field. Due to the dephasing during this precession, the spin accumulation is gradually reduced with an approximately Lorentzian shape.[3,4] The extra voltage at the interface associated to the spin accumulation (spin voltage) can be written as:

$$\Delta V(B_\perp) = \frac{\Delta V_{HANLE}}{1+(\omega_L \tau_{SF})^2} . \qquad (1)$$

$\Delta V_{HANLE}$ is the value of spin voltage in the absence of $B_\perp$, $\tau_{SF}$ the spin relaxation time, and $\omega_L = g\mu_B B_\perp/\hbar$ is the Larmor frequency, where $g$ is the Landé g-factor, $\mu_B$ is the Bohr magneton and $\hbar$ is the reduced Planck's constant. Commonly, the spin voltage is normalized to the injected current and, thus, $\Delta R_{HANLE} = \Delta V_{HANLE}/I$. In Fig. 2 (solid symbols) we show two examples of a Hanle-like curve measured at 10 K, obtained after subtracting the quadratic background from the data.[7] Fig. 2(a) corresponds to Sample 1 (measurements done at I=-100 µA), whereas Fig. 2(b) shows the same experiment on Sample 2 (I=-5 µA). We should also observe a change in the measured voltage if we apply an in-plane magnetic field ($B_{II}$).[11] It has been reported that, due to the roughness of the interface, local magnetic fields appear on the non-magnetic material and make the injected spins precess even in the absence of $B_\perp$. Therefore, the precession is suppressed by applying $B_{II}$ and the spin accumulation increases until it saturates for large enough fields.[11] This is the so-called inverted Hanle effect, with associated signal $\Delta R_{INV}$. The data measured at 10 K on Sample 1 (performed at I=-100 µA) and Sample 2 (I=-5 µA) are represented by the open symbols in Fig. 2(a) and (b), respectively. According to Ref. 11, the total spin accumulation is proportional to the difference between the saturation value in the inverted Hanle effect curve and the minimum value of the Hanle effect curve. We call this difference the total Hanle signal, $\Delta R_{TOT} = \Delta R_{HANLE} + \Delta R_{INV}$. At 10 K and previously described biases, $\Delta R_{HANLE}$ values are 26% and 21% of $\Delta R_{TOT}$ for Sample 1 and 2, respectively. These percentages are in agreement with those reported in Ref. 11. Fig. 2 shows that very similar results are obtained for both samples, suggesting that properties of the tunnel oxide barrier are not relevant.

The standard theory of spin injection and accumulation in the diffusive regime[28,29] states that the total spin signal associated to spin accumulation at the interface is given by $\Delta R_{TOT} = \gamma^2 R_N$, in the limit $R_{int} \gg R_N \gg R_F$, which is our case. $\gamma$ is the tunneling spin polarization and $R_{N(F)} = \rho_{N(F)} \lambda_{N(F)}^2 / V_{N(F)}^S$ is the spin resistance of the non-magnetic (ferromagnetic) side of the interface, where $\rho_{N(F)}$ are the resistivities, $\lambda_{N(F)}$ the spin diffusion lengths and $V_{N(F)}^S$ the effective volumes of spin relaxation.[30] Assuming that the spin injection occurs homogeneously along all the contact area and that, for the device sketched in Fig. 1(a), $w_F, w_N \gg \lambda \gg d$ is satisfied, then $V_N^S = w_F \cdot w_N \cdot d$. Therefore, the total spin signal can be expressed as:

$$\Delta R_{TOT} = \frac{\gamma^2 \rho_N \lambda_N^2}{w_F w_N d} . \qquad (2)$$

Taking into account the $\lambda_N$ values obtained from the fitting to Eq. 1 (see below) and that typical γ values for Py/Al$_2$O$_3$ interfaces are 0.02-0.25,[18,19] the total spin signal should be of the order of $10^{-8}$ Ω, whereas the measured values are at least 8 orders of magnitude higher (see Fig. 2).

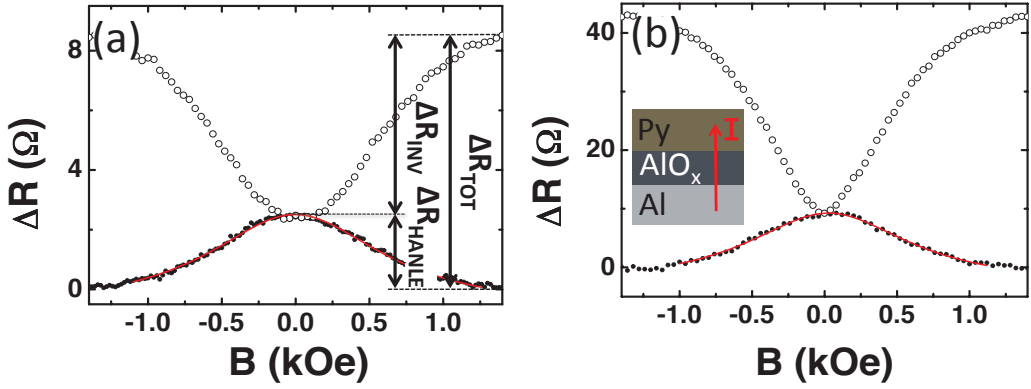

FIG. 2. (a) Hanle and inverted Hanle-like curves (solid and open circles, respectively) for Sample 1, measured at 10 K and -100 μA. Red solid line is the Lorentzian fit of the data to Eq. 1. Hanle ($\Delta R_{HANLE}$), inverted Hanle ($\Delta R_{INV}$) and total Hanle ($\Delta R_{TOT}$) signals are defined. (b) Same as (a) for Sample 2, measured at 10 K and -5 μA. The device scheme is shown as an inset.

Discrepancies between theoretical and experimental values of spin signal have previously been observed and discussed for semiconductors.[3,5-11] The small theoretical spin signal could be caused by the decrease of $\tau_{SF}$ due to the broadening of the Hanle curve,[3,11] although, in our case, this effect is by far not enough to explain the discrepancy. The role of localized states near the tunnel junction/semiconductor interface has also been deeply debated lately.[3,5-11] In our samples, since Al is a conductor, such states could only be created within the oxide tunnel junction due to fabrication conditions. However, evidences of the existence of localized states, such as an strong temperature dependence of $R_{int} \cdot A$, or a barrier height $\phi$ larger than expected are not found,[31] neither in Sample 1, with the tunnel barrier grown by plasma, nor in sample 2, grown by natural oxidation. Finally, Dash et al.[3] analyze the possibility of having lateral inhomogeneities at the tunnel junction. In this case, electrons mostly tunnel through the thinnest regions of the junction, so-called hot spots. This scenario is probable in our tunnel barriers due to the inherent roughness of the AlO$_x$ surface (r.m.s.=0.7 nm for Sample 1 and 0.4 nm for Sample 2). In the presence of hot spots, the effective volume of spin accumulation would be reduced, leading to an enhancement of the theoretical spin signal. We can recalculate $V_N^S$ by assuming the existence of N hot spots on the tunnel barrier. If the size of these spots is smaller than $\lambda_N$, and the distance between them is longer than $2\lambda_N$, then $\Delta R_{TOT} = \gamma^2 \rho_N/N\pi d$. In the limiting unrealistic case with N=1, which gives the smallest effective volume, we find that $\Delta R_{TOT} \approx 10^{-2}$ Ω, and the theoretical Hanle signal is still two orders of magnitude lower than the experimental one. Therefore, the enormous difference between standard theory and experiments cannot be explained by the existence of hot spots.

Next, we show that the temperature and bias dependencies of the total Hanle signal cannot be explained on the basis of Eq. 2. Fig. 3(a) shows $\Delta R_{TOT}$ as a function of temperature both for Sample 1 and 2. Since the spin diffusion length is inversely proportional to the resistivity,[21] then $\Delta R_{TOT} \propto \gamma^2 \rho_N^{-1}$, where $\gamma$ can be expressed as[32] $\gamma = \gamma_0(1 - (T/T_C)^{3/2})$, being $T_C$ the Curie temperature of the ferromagnet. Using this expression for $\gamma$ and the experimental values for $\rho_N^{-1}$, we fitted $\Delta R_{TOT}(T)$ to extract $T_C$ for Sample 1 and 2 (Fig. 3(a)). For Sample 1, we obtain $T_C = (516 \pm 13)\ K$, which could be in agreement with literature values for Py.[33] However, for Sample 2, we obtain $T_C = (159 \pm 7)\ K$, in disagreement with a tunneling spin injection from Py. Fig. 3(b) shows the voltage-dependent measurements of $\Delta R_{TOT}$ at 10 K. For both Sample 1 and 2, we observe that the signal becomes undetectable at low bias voltage ($|V| \leqslant 0.5\ V$ in Sample 1 and $|V| \leqslant 0.025\ V$ in Sample 2). This gap in $\Delta R_{TOT}(V)$ at low bias cannot be explained by the standard theory of spin injection. Indeed, the tunneling spin polarization $\gamma$ is the only bias-dependent parameter in Eq. 2, and it is largest at low bias.[20,34]

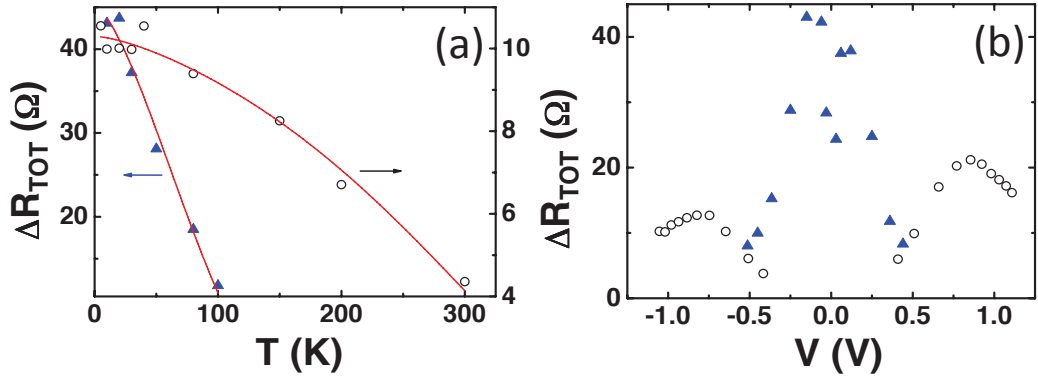

FIG. 3. (a) Total Hanle signal as a function of temperature, measured at -100 µA for Sample 1 (open circles) and at -5 µA for Sample 2 (solid triangles). Red solid lines are fits of the data to Eq. 2. (b) Total Hanle signal as a function of the applied bias at 10 K, for Sample 1 (open circles) and Sample 2 (solid triangles).

Concerning the spin relaxation time, the obtained Lorentzian curves have been fitted to Eq. 1 to extract the value for Al: in the case of Sample 1, we obtain $\tau_{SF} = (80 \pm 2)$ ps at 10 K and $(73 \pm 9)$ ps at 300 K. For Sample 2, the Lorentzian curve vanishes above 100 K, obtaining $\tau_{SF} = (82 \pm 3)$ ps at 10 K and $(43 \pm 13)$ ps at 100 K. These values are the same for all biases (including injection into Al and extraction from Al). From $\tau_{SF}$, spin diffusion length values $\lambda_{Al}$ can be calculated as $\lambda_{Al} = \sqrt{D\tau_{SF}}$, where $D$ is the diffusion coefficient. The obtained values in Sample 1 are $\lambda_{Al} = (439 \pm 5)$ nm at 10 K and $(370 \pm 20)$ nm at 300 K, whereas in Sample 2 we obtain $\lambda_{Al} = (325 \pm 6)$ nm at 10 K and $(232 \pm 35)$ nm at 100 K. Therefore, while there is no agreement neither with the amplitude, nor with temperature dependence and bias dependence of the accumulation, the spin diffusion length values obtained in the two samples seem to be in agreement with literature.[4,17-21]

In order to clarify this controversy, similar measurements are performed in Au, another metal whose spin transport properties are also well-known[21-24] but are very different to the ones in Al. Fig. 4 shows the Hanle and inverted Hanle-like curves measured at 10 K for I=-50 µA in Sample 4. The spin relaxation time value extracted from the fitting of the Lorentzian curve (red solid line in Fig. 4) is (144±5) ps, which is two orders of magnitude higher than the expected values for Au ($\tau_{SF} \sim$ 1 ps).[21-24] Moreover, total Hanle amplitudes above the theoretically expected values (see Fig. 4) and anomalous voltage-dependence of signal amplitude (not shown) have also been measured in this sample. Although the spin relaxation times obtained for Al (Samples 1 and 2) are reasonable,[4,17-21] similarities between the results reported in Al and Au 3T devices (comparable spin relaxation time, amplitude of signals above the expected values, and anomalous bias dependence of the total Hanle signal) evidence that the measured Hanle and inverted Hanle-like curves are not originated by spin accumulation in the metal.

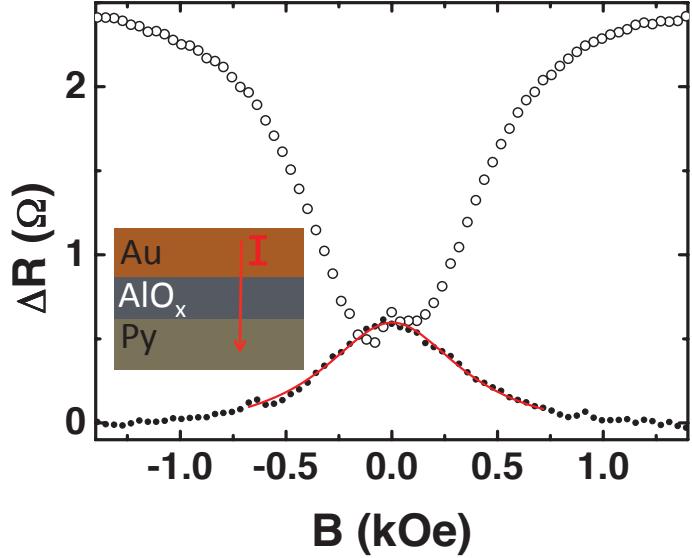

FIG. 4. Hanle and inverted Hanle-like curves (solid and open circles, respectively) measured in Sample 4 at 10 K and -50 µA. Red solid line is the Lorentzian fit of the data to Eq. 1. The device scheme is shown as an inset.

Fig. 5 shows the spin RA product ($\Delta R_{TOT} \cdot A$) as a function of the RA product of the tunnel junctions ($R_{int} \cdot A$) of all 3T devices used in this work, which employ different metals and tunnel barriers obtained through different fabrication processes. A clear scaling between $\Delta R_{TOT} \cdot A$ and $R_{int} \cdot A$, with a power law exponent of 1.31, is observed in our Py/AlO$_x$/metal devices with different RA products. This result, which spans over 2 orders of magnitude, is not predicted by the standard theory in the condition $R_{int} \gg R_N \gg R_F$.[28,29] The scaling suggests again that the Hanle and inverted Hanle-like curves do not arise from the modulation of the spin accumulation in the metal by an external magnetic field, but from the tunnel junction itself. Recent reports on 3T measurements in Si- and Ge-based devices have shown similar disagreements. Aoki et al.[14] report a Lorentzian curve in both Hanle and inverted Hanle configuration, with a corresponding spin relaxation time of ~50 ps, which cannot be associated to spin accumulation in Si when compared to non-local four-terminal measurements. Uemura et al.[15] and Sharma et al.[35] report a tunnel barrier thickness dependence of Hanle signals, observing a scaling of the spin RA product with the RA product of the junction as well. It is worth noting that the spin relaxation time obtained in Ref. 15 for n-Si is 150 ps, very close to our values both for Al and Au.

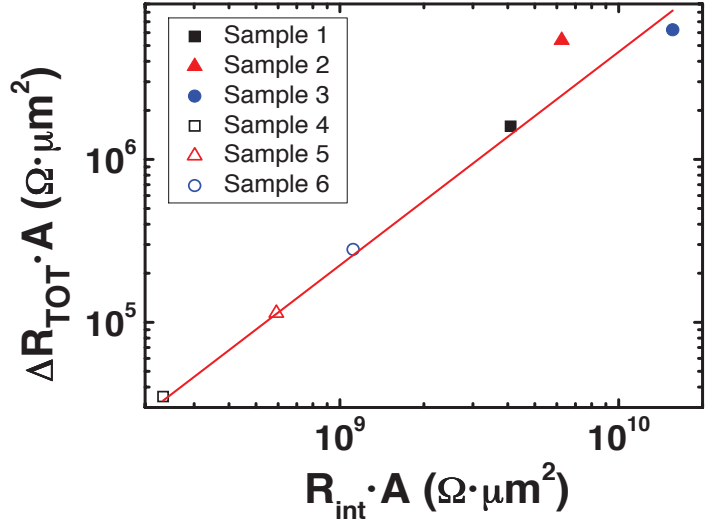

FIG. 5. Spin RA product as a function of the RA product of the tunnel junction, for different samples measured at 10 K and their optimum bias conditions. Solid and open symbols are for Al and Au samples, respectively. Red line is the logarithmic fit to the data.

In conclusion, we have tested the reliability of three-terminal Hanle measurements in metals such as Al and Au, with well-known and different spin transport properties. Our results indicate that the obtained Hanle and inverted Hanle-like curves are not related to spin accumulation in the metal channel. First, and most important, the spin relaxation times obtained for both metals are comparable, even if much shorter values are expected for Au. Furthermore, these values are also comparable to those reported in other systems such as semiconductors using the same configuration.[14,15] Second, the total Hanle signal is several orders of magnitude higher than the value predicted from the standard theory. Such a difference cannot be explained by any of the sources previously discussed in literature.[3,5-11] Third, the temperature and voltage dependences of the signal are not in agreement with the tunneling spin polarization of the used ferromagnetic contact. Last, we show that the total Hanle signal is proportional to the interface resistance of the tunnel barrier, a dependence which cannot be explained by models proposed so far and which suggests that the observed effect originates at the tunnel barrier.

Finally, we would like to emphasize that great care should be taken (e.g., use of control samples) when measuring the Hanle effect using three-terminal geometries with highly resistive tunnel barriers to extract spin injection and transport properties from any type of materials, including metals and semiconductors.

This work is supported by the European Union 7th Framework Programme (NMP3-SL-2011-263104-HINTS, PIRG06-GA-2009-256470 and the European Research Council Grant 257654-SPINTROS), by the Spanish Ministry of Science and Education under Project No. MAT2009-08494 and by the Basque Government under Project No. PI2011-1. We would also like to acknowledge Prof. Casey W. Miller for fruitful discussions.